\documentclass[10pt,conference,onecolumn]{IEEEtran}

\IEEEoverridecommandlockouts

\usepackage{adjustbox}
\usepackage{subcaption}
\usepackage{caption}
\usepackage{graphicx} 
\usepackage{stfloats}
\usepackage{booktabs}
\usepackage{multirow}
\usepackage[percent]{overpic}
\usepackage{graphicx}
\usepackage{amsmath,amssymb,amsfonts}
\usepackage{siunitx}
\usepackage{float}
\usepackage{multicol}
\usepackage{lipsum} 
\usepackage{textcomp}
\usepackage{xcolor}
\usepackage{breakurl}
\usepackage{xurl}
\usepackage{longtable} 
\usepackage{array}     
\usepackage{verbatim}
\usepackage[ruled, vlined, linesnumbered, boxed]{algorithm2e}
\usepackage{algpseudocode}

\title{Detection of Misreporting Attacks on Software-Defined Immersive Environments\thanks{This material is based upon work supported by the National Science Foundation (NSF) under Award Number CNS-2401928.}}
\author{
Sourya Saha, Md. Nurul Absur, Shima Yousefi, Saptarshi Debroy\\
City University of New York\\ 
Emails: \textit{\{ssaha2,mabsur,syousefi\}@gradcenter.cuny.edu, saptarshi.debroy@hunter.cuny.edu}
}
\date{May 2025}

\begin{document}

\maketitle
\thispagestyle{empty}
\pagestyle{empty}
\maketitle

\begin{abstract}
The ability to centrally control network infrastructure using a programmable middleware has made Software-Defined Networking (SDN) ideal for emerging applications, such as immersive environments. 
However, such flexibility introduces new vulnerabilities, such as switch misreporting led load imbalance, which in turn make such immersive environment vulnerable to severe quality degradation.  
In this paper, we present a hybrid machine learning (ML)-based network anomaly detection framework that identifies such stealthy misreporting by capturing temporal inconsistencies in switch-reported loads, and thereby counter potentially catastrophic quality degradation of hosted immersive application. The detection system combines unsupervised anomaly scoring with supervised classification to robustly distinguish malicious behavior. 
Data collected from a realistic testbed deployment under both benign and adversarial conditions is used to train and evaluate the model. Experimental results show that the framework achieves high recall in detecting misreporting behavior, making it effective for early and reliable detection in SDN environments.
\end{abstract}


\section{Introduction}

Software-Defined Networking (SDN) has become a key technology for next-generation networks due to its centralized control, real-time programmability, and dynamic flow management. Such flexibility enables intelligent orchestration of resources, making SDN attractive for performance-critical applications. Immersive 3D applications, such as Virtual and Augmented Reality (VR/AR), is a prominent domain which can leverage SDN to enhance situational awareness in areas such as cyber-training, healthcare, and emergency response~\cite{sec2023}. For instance, SDN has been applied to improve VR streaming through MCTS-based routing \cite{mcts}, multipath delivery for tiled content \cite{multipath}, and ML-driven network slicing for latency-sensitive VR \cite{6gsdnarvr}.



However, SDN’s centralized design and hardware-software coupling create a broad attack surface. Known attacks include flow table overflow, packet injection, and traffic flooding, which can destabilize both SDN systems and hosted critical applications. A particularly subtle vulnerability is \emph{misreporting}, wherein a malicious switch(s) falsifies load statistics to bias the controller’s flow assignments. This leads to unfair traffic distribution and performance degradation in latency-sensitive workloads. In VR, such manipulation can degrade quality of experience (QoE) by redirecting tasks to compromised servers. 
Such manipulated pose updates may lead to visually inconsistent rendering, emphasizing the need for timely detection.

While defense mechanims, such as rDefender \cite{rdefender}, counter-flow \cite{ddos-sdn}, and SDN-Guard \cite{sdn-guard} address rule-based vulnerabilities, effective strategies against switch misreporting attacks are fewer to almost none. Detection of such attacks is challenging because misreported statistics mimic historical patterns, evading threshold-based or anomaly filters. Recent work \cite{misreporting} shows how compromised switches can attract traffic without triggering alarms, exposing a gap in current SDN defenses that largely assume a trusted data plane.


In this paper, we study stealthy misreporting in SDN and its impact on latency-sensitive applications. Using VR offloading as a case study, we deploy our setup on the NSF FABRIC testbed~\cite{fabric-2019}, replicating the misreporting model of \cite{misreporting} to show how falsified load reports can redirect workflows to a malicious edge server that perturbs VR pose updates. Integrated with the ILLIXR environment~\cite{illixr}, our setup traces the attack’s effect from network telemetry to application-level QoE. To counter this, we propose a hybrid detection framework combining statistical features with temporal modeling to reveal anomalies that appear plausible individually but suspicious over time. A transformer-based autoencoder trained on benign telemetry yields unsupervised anomaly signals—reconstruction error, Mahalanobis distance, and rolling-window statistics (z-score, skewness, kurtosis). These signals are then fused with lightweight supervised models—an MLP and calibrated LightGBM—balancing generalization with precision in detecting subtle misreporting.


We evaluate the proposed framework on an SDN testbed built on FABRIC with controlled misreporting. ILLIXR environment serves as a VR layer, offloading head pose estimation to edge servers hosted on FABRIC, letting us measure QoE degradation when workflows are steered to a compromised server. The setup demonstrates higher Absolute Trajectory Error (ATE) and Relative Pose Error (RPE), confirming impact on user experience. Benchmarks, such as precision, recall, F1, and ROC AUC, show up to 20\% F1 gain over unsupervised baselines with latency suitable for real-time deployment. System robustness is also assessed under varying window sizes, strides, VR durations, and attack persistence, providing early evidence of effective misreporting detection in SDN-hosted immersive environments.

The remainder of this paper is organized as follows. Section \ref{sec:rel-works} reviews relevant prior work. 
Section \ref{sec:sys-threat-model} presents the system and threat models.
Section \ref{sec:detection} outlines the proposed detection framework.
Section \ref{sec:eval-results} presents our evaluation methodology and experimental results. Section \ref{sec:Conclusion} concludes the paper and discusses future work.


\section{Related Works}
\label{sec:rel-works}

Falsification-based threats remain a major challenge in SDN and cyber-physical systems (CPS), where accurate reporting from distributed components is critical. These attacks exploit trust in telemetry—whether switch statistics, sensor readings, or link metrics—to mislead controllers or higher-level logic. In SDN, the consequences are especially severe given centralized control. The Marionette attack \cite{marionette} shows how high-priority flow entries can redirect LLDP packets and fabricate a fake but plausible topology. Similarly, the Link Latency Attack (LLA) \cite{link-layer-latency} manipulates latency estimates through ARP flooding and LLDP relaying. Such examples highlight the fragility of data-driven decision-making in programmable networks.


Within this broader class of falsification attacks, misreporting attacks at the SDN data plane are particularly insidious: compromised switches falsify traffic statistics to deceive the controller. Our work builds on \cite{misreporting}, which introduced a trivial zero-reporting attack and a stealthier variant drawing fake values from historical distributions. Both significantly skew load balancing, with misreporting switches attracting over 200\% more traffic while staying within a 2–10\% deviation. Further analysis in \cite{stealthy} shows such attacks can be tuned via reconnaissance and achieve provable stealth under bounded perturbations, bypassing threshold-based anomaly detectors.


Misreporting led deception also appears in other domains, where diverse defenses have been proposed. SPHINX \cite{sphinx} offers an FSM-based model that captures expected SDN control plane behavior and detects deviations caused by unauthorized rule insertions or topology manipulations. Statistical approaches include Kalman filters \cite{dynamic-load-altering}, z-score filtering \cite{data-driven}, and autoencoders for CPS anomaly detection \cite{cps-based}. More advanced methods apply GANs \cite{gan-sdn}, explainable ensembles \cite{ensemble-unknown}, and hybrid statistical-ML designs \cite{quantum-misrpeorting}. {\em While promising, most assume adversarial data lies outside historical distributions. Stealthy misreporting instead mimics legitimate behavior, evading outlier-based detection.}

In centrally controlled VR systems, the implications are pronounced. VR requires ultra-low latency and high throughput to sustain immersion, with motion-to-photon (MTP) latency—the time between a user’s movement and corresponding visual feedback—typically constrained below 20 ms \cite{bassbouss2016high} to avoid dizziness and disorientation. To meet this, compute-intensive tasks such as Visual-Inertial Odometry (VIO) or rendering are increasingly offloaded to edge servers \cite{remotevio}. This creates dependencies on SDN-based routing and switch telemetry, where controller decisions determine task placement. Recent work like XRgo \cite{xrgo} and RemoteVIO \cite{remotevio} show clear performance benefits in power and stability. {\em Yet none of these systems consider adversarial SDN behavior: a misreporting switch could bias selection toward a compromised server, injecting subtle pose perturbations or spatial drift while still meeting latency budgets—producing degradations invisible to conventional QoS metrics. Although multipath and ML-driven optimizations for VR/AR traffic have been explored \cite{mcts, multipath, 6gsdnarvr}, the risks of stealthy misreporting in latency-sensitive immersive applications remain underexplored.}


\begin{figure}[t]
  \centering
  \includegraphics[width=0.95\columnwidth]{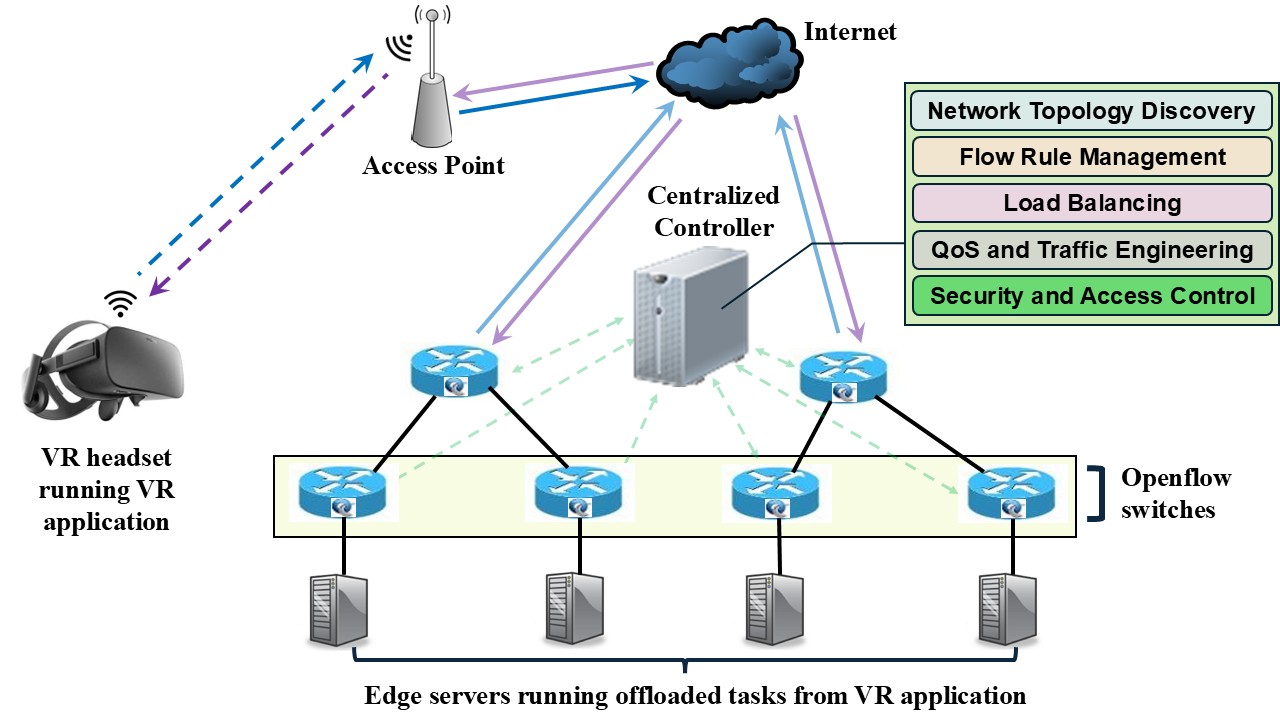}
  \caption{VR pipeline offloading to remote edge servers using SDN infrastructure}
  \label{fig:sdn-setup}
  \vspace{-0.1in}
\end{figure}


\section{System and Threat Model}
\label{sec:sys-threat-model}


\subsection{The application pipeline}





We consider a VR system deployed over an SDN infrastructure, where compute-intensive components such as rendering or pose estimation are offloaded from the client to remote edge servers. The system runs in fixed-length \textit{workflows} (e.g., 15 seconds), each dynamically routed by the SDN controller based on real-time network conditions. As shown in Fig.~\ref{fig:sdn-setup}, the VR client initiates workflows and exchanges data with selected servers through a wireless access network. Edge servers perform the offloaded tasks and return results via SDN-managed switches. These OpenFlow switches forward packets and report per-port statistics (e.g., byte counters) that guide controller decisions. The controller, beyond standard functions, polls switches periodically and assigns each new workflow to the server behind the switch reporting the lowest load, aiming to balance traffic and prevent congestion.


\subsection{Misreporting attack model}

While this architecture enables efficient offloading, it relies on trusted switch statistics. An attacker who compromises a server and its switch can exploit this trust to degrade VR QoE. Attack vectors (e.g., malware, Trojans) are beyond our scope. Once compromised, the attacker falsifies port-level load statistics~\cite{misreporting} to influence controller decisions, redirecting workflows. Subtle pose manipulations then accumulate across routing intervals, reducing visual coherence and overall VR experience.





We adopt the stealthy misreporting model of \cite{misreporting} as a reproducible adversarial mechanism to stress-test detection. The SDN controller polls edge switches for port-level byte counts and assigns workflows to the switch reporting the lowest load. A compromised switch falsifies reports, with probability \(\varphi\), replacing its true load with a value from the bottom \(\rho\)-th percentile of its historical loads—blending into natural traffic variation. To attract a target fraction \(\tau\) of workflows, the attacker sets:

\begin{equation}
\label{eq:phi}
\scriptsize
\varphi = \frac{\tau - \frac{1}{S}}{(1 - \rho)^{S - 1} - \frac{1}{S}}
\end{equation}

where \(S\) is the number of switches and \(\rho\) controls stealthiness. The attack persists for a bounded epoch window \(\epsilon\), after which misreporting stops.

\begin{figure*}[htbp]
\centering
\includegraphics[width=\textwidth]{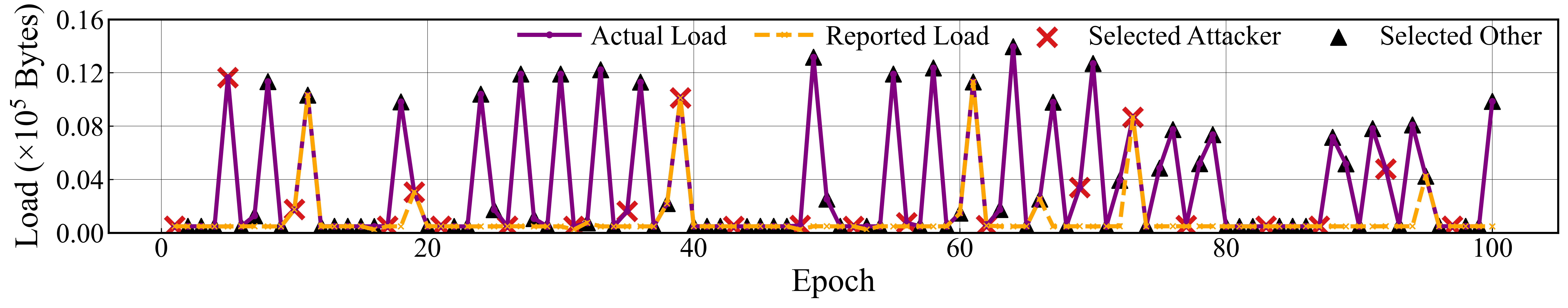}
\vspace{-0.2in}
\caption{Reported and actual load values across 100 epochs during a misreporting interval} 
\label{fig:misreporting_visual2}

\end{figure*}

Fig.~\ref{fig:misreporting_visual2} shows a 100-epoch window from our dataset with \(\tau=0.6\), \(\epsilon=1000\), and \(\rho=0.01\). It compares actual and reported loads at the compromised switch alongside controller decisions. Misreporting occurs when reported load falls below the true load, biasing selection. Red ``\(\times\)'' markers denote epochs where the compromised switch was chosen, and black ``\(\triangle\)'' markers indicate other switches. The results show misreporting raises the compromised switch’s selection probability while preserving stealth.


Our goal is to design a detection mechanism that identifies malicious switches early in the workflow sequence, preventing further compromise. Mitigation is beyond this paper’s scope, though a possible response is provided in Section~\ref{sec:Conclusion}.


\subsection{Quantitative Evaluation of QoE Degradation Under Misreporting: A VR Case Study}

To motivate detection, we quantify misreporting’s impact on VR QoE. While it may seem counterintuitive that falsified switch statistics alone degrade QoE, our experiments confirm this. Repeated redirection of pose-estimation tasks to a compromised server—via stealthy misreporting—lets the attacker inject controlled noise into pose values before returning them to the client, similar to \cite{saha2025detectionrecoveryadversarialslowpose}. These perturbations disrupt the motion-to-photon pipeline, reducing visual coherence and introducing spatial inconsistencies that degrade immersion.


Unlike the Mininet-based setup in \cite{misreporting}, our deployment uses edge servers on the Internet-scale FABRIC testbed. To handle higher control-plane latency, we adopt a 2-second polling interval (vs.~1s) for reliable statistics without reducing attack effectiveness. The switching fabric runs Open vSwitch on FABRIC nodes, instrumented with a modified library supporting \emph{normal} and \emph{attack} modes. In normal mode, switches return true per-port byte counters via \texttt{dump-ports}; in attack mode, each report follows a Bernoulli trial with frequency \(\varphi\), and falsified loads are sampled from the bottom \(\rho\)-th percentile of historical values to mimic plausible traffic. Attacker parameters—\(\tau\) (target share), \(\rho\) (stealth), \(\epsilon\) (window), and \(S\) (switch count)—are coded in, allowing \(\varphi\) to be computed via Equation~\ref{eq:phi} and tuned manually. Redirected workflows trigger ILLIXR server modifications that inject noise into pose updates, degrading QoE of client-side VR. This setup forms the adversarial baseline for detection. Impact is measured by comparing estimated trajectories against ground truth using standard error metrics~\cite{rpeate}:

\begin{itemize}
    \item \textbf{Absolute Trajectory Error (ATE)}: Evaluates global consistency of predicted poses. After aligning predictions to ground truth via a rigid-body transform, ATE is the RMSE of differences in corresponding poses:

    \begin{equation}
    \scriptsize
    \mathbf{e}_i = \hat{\mathbf{T}}_i^{-1} \, \mathbf{S} \, \mathbf{T}_i
    \end{equation}

    \begin{equation}
    \scriptsize
    \text{ATE}_{\text{RMSE}} = \sqrt{ \frac{1}{N} \sum_{i=1}^{N} \left\| \text{trans}(\mathbf{e}_i) \right\|^2 }
    \end{equation}

    where \( \mathbf{T}_i \) is the estimated pose, \( \hat{\mathbf{T}}_i \) the ground truth, \( \mathbf{S} \) the optimal alignment using Horn's method, and \( N \) the number of poses. The translational component of the pose error \( \mathbf{e}_i \) is extracted using \( \text{trans}(\cdot) \).

    \item \textbf{Relative Pose Error (RPE)}: Captures local motion fidelity via RMSE of relative pose differences over fixed time interval \(\delta\):

    \begin{equation}
    \scriptsize
    \mathbf{r}_i := \left( \hat{\mathbf{T}}_i^{-1} \hat{\mathbf{T}}_{i+\delta} \right)^{-1} \left( \mathbf{T}_i^{-1} \mathbf{T}_{i+\delta} \right)
    \end{equation}

    \begin{equation}
    \scriptsize
    \text{RPE}_{\text{RMSE}}(\delta) = \sqrt{ \frac{1}{M} \sum_{i=1}^{M} \left\| \text{trans}(\mathbf{r}_i) \right\|^2 }
    \end{equation}

    where \( M \) is the number of intervals and \( \text{trans}(\cdot) \) extracts translation.
\end{itemize}


To evaluate the impact of compromised pose estimation, we simulate an adversary that intermittently alters edge server pose outputs. Spoofing is applied at four levels: 0\% (none), 25\% (every fourth pose altered), 50\% (alternate poses altered), and 75\% (three of four poses altered). These perturbations reduce pose accuracy and rendering quality, producing artifacts such as jitter, reduced immersion, and VR fatigue.

\begin{figure}[t]
    \centering
    \includegraphics[width=0.7\columnwidth]{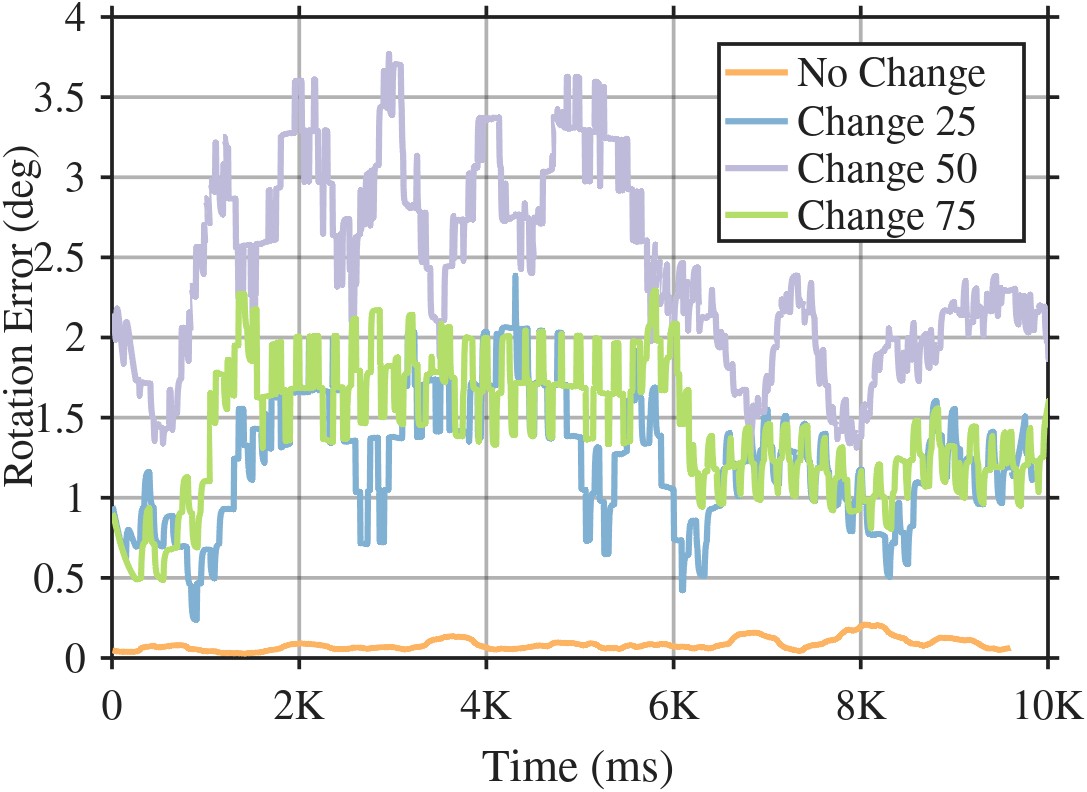}
    \caption{Smoothed ($\text{window}=100$) RPE rotation error over time for different levels of pose spoofing}
    \label{fig:rpe_rotation_all}
    \vspace{-0.1in}
\end{figure}

Fig.~\ref{fig:rpe_rotation_all} shows rotational RPE. At 0\%, errors stay flat, indicating stable tracking. At 25\%, periodic deviations emerge; at 50\%, alternating real and corrupted poses cause sharp oscillations and perceptual jitter. At 75\%, errors are smoother but consistently high, reflecting orientation drift. Thus, mid-frequency spoofing (50\%) yields greater instability than occasional (25\%) or continuous (75\%) corruption.

\begin{figure}[t]
    \centering
    \begin{subfigure}[t]{0.49\linewidth}
        \centering
        \includegraphics[width=\linewidth]{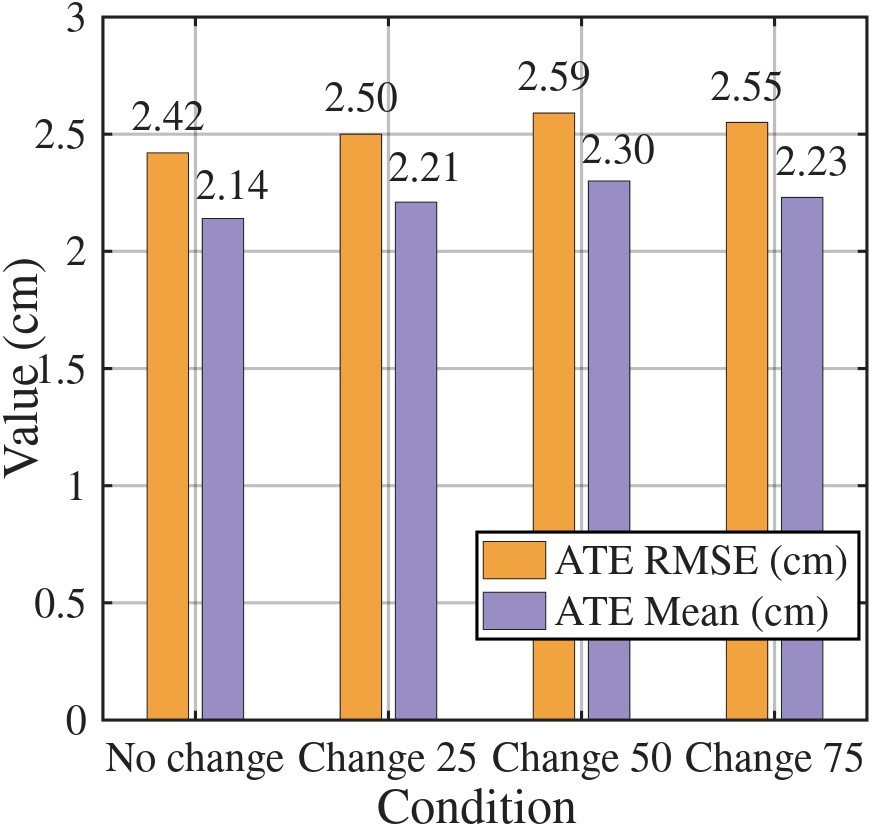}
        \caption{ATE RMSE and mean for client-side trajectories.}
        \label{fig:ate-summary}
    \end{subfigure}
    \hfill
    \begin{subfigure}[t]{0.49\linewidth}
        \centering
        \includegraphics[width=\linewidth]{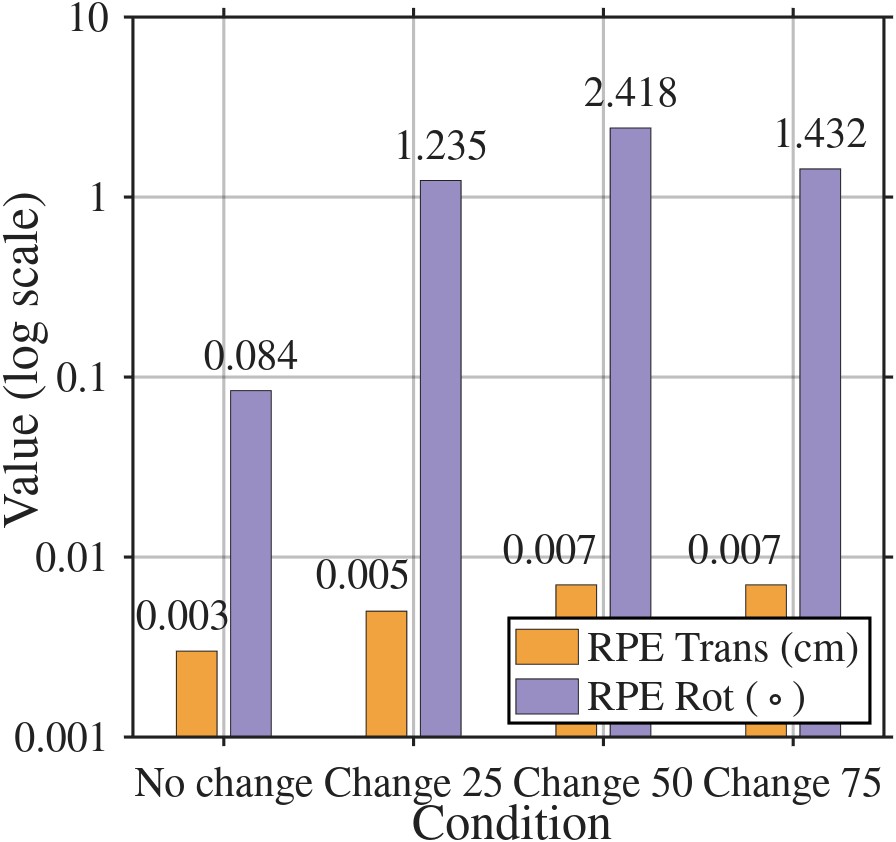}
        \caption{Log-scale RPE translation and rotation errors.}
        \label{fig:RPEerrorlog}
    \end{subfigure}
    \vspace{-1mm}
    \caption{Evaluation metrics under edge perturbation. (a) Trajectory accuracy (ATE). (b) Pose drift behavior (RPE).}
    \label{fig:perturbation-summary}
    \vspace{-0.1in}
\end{figure}


Fig.~\ref{fig:ate-summary} shows ATE increasing with spoofing, with 50\% yielding the largest trajectory error. Frequent toggling between correct and corrupted poses accumulates misalignment, showing that inconsistency—not just intensity—harms coherence. Fig.~\ref{fig:RPEerrorlog} presents translational RPE (log scale): errors rise with spoofing, with 50\% and 75\% reaching similar levels, while rotational RPE peaks at 50\%. Overall, mid-frequency spoofing causes the greatest rotational disruption, and higher spoofing worsens translational accuracy.


\section{Detection Model}
\label{sec:detection}


To illustrate detection challenges, we show how stealthy misreporting biases controller decisions while evading simple threshold filters. The attacker substitutes true byte counts with values from the lower tail of its history, which appear plausible in isolation. As Fig.~\ref{fig:misreporting_visual2} shows, with $\tau=0.60$, $\rho=0.01$, and $\varphi \approx 0.48$, misreporting can dominate most epochs yet remain statistically believable. This underscores the need for models that exploit temporal patterns and switch-specific baselines rather than point-wise thresholds.


ML offers tools to capture such subtle behaviors across domains. In healthcare, hybrid methods combining isolation forests with supervised classifiers improved anomaly detection in electronic records~\cite{tabassum2024anomaly}. In autonomous driving, LIFE exploited sensor correlations to detect spoofing~\cite{liu2021seeing}, while smart grid defenses used autoencoder–GAN hybrids to classify attacks~\cite{siniosoglou2021unified}. IoT systems applied fog-based adaptive learning for real-time detection~\cite{hameed2021hybrid}.


Motivated by these works, we design a hybrid ML framework for misreporting detection in SDN. Since only reported switch values are observable and Fig.~\ref{fig:misreporting_visual2} shows no per-point distinction between real and fake reports, we extract temporal features to amplify deviations. Features span four groups: \textit{basic load statistics} (raw load, deltas, rolling mean); \textit{distributional indicators} (percentiles, z-score, skewness, kurtosis); \textit{temporal stability} (recent standard deviation, autocorrelation, Mean Absolute Deviation); and \textit{peer context} (load ratios, mean delta across peers, unique counts within a window).

\begin{figure}[t]
\centering
\includegraphics[width=0.43\textwidth]{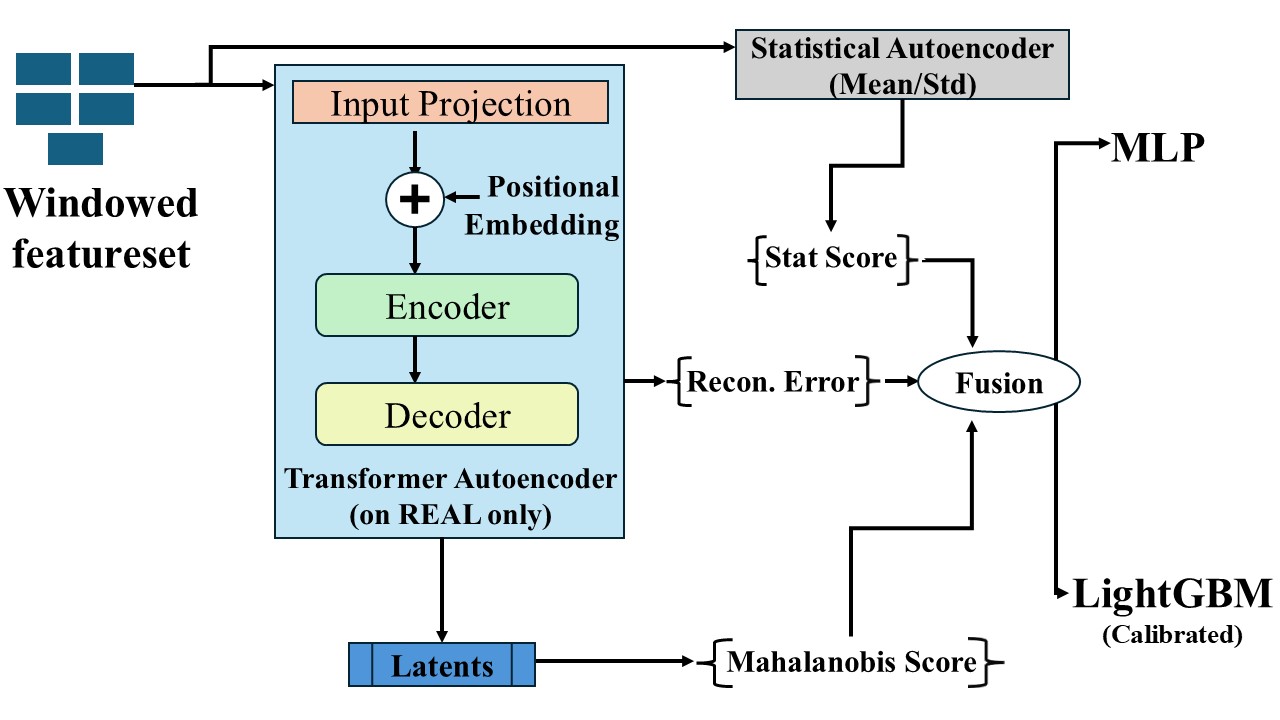}
\caption{Pipeline of hybrid detection model}
\label{fig:ml-pipeline}
\vspace{-0.1in}
\end{figure}


Features are computed per switch over overlapping sliding windows. A window is labeled FAKE if any timestep is misreported, otherwise REAL, enabling detection of localized anomalies. This representation preserves temporal continuity and supports sequential models such as Transformers or LSTMs. As shown in Fig.~\ref{fig:ml-pipeline}, our pipeline uses a Transformer autoencoder trained on REAL windows to capture benign dynamics and output two unsupervised cues: reconstruction error and Mahalanobis distance. In parallel, a statistical autoencoder produces deviation scores against rolling statistics. These three signals—reconstruction, Mahalanobis, and statistical—form the fused anomaly feature triplet.


Scores are refined using lightweight classifiers—a one-layer MLP and a calibrated LightGBM. LightGBM is calibrated on the validation set with three fused features (reconstruction error, statistical deviation, Mahalanobis distance) via 5-fold cross-validation and sigmoid scaling. This hybrid design adapts to varied attack profiles and improves detection fidelity. Tab.~\ref{tab:transformer_thresholds} highlights the limitation of relying only on the Transformer autoencoder: reconstruction error achieves high F1 and accuracy for \textit{REAL} samples, but F1$_\text{FAKE}$ falls from 0.71 at the 90th percentile to <0.5 at the 96th as thresholds tighten. This decline shows stealthy misreporting evades detection under reconstruction loss alone, motivating our hybrid approach that augments latent-space (Mahalanobis) and statistical deviation cues with supervised refinement.


\begin{table}[htbp]
    \centering
    \footnotesize
    \caption{Transformer autoencoder performance}
    \label{tab:transformer_thresholds}
    \resizebox{\linewidth}{!}{%
    \begin{tabular}{
        S[table-format=2.0]
        S[table-format=1.6]
        S[table-format=1.4]
        S[table-format=1.4]
        S[table-format=1.4]
        S[table-format=1.4]
    }
        \toprule
        \multicolumn{1}{c}{\textbf{Percentile}} &
        \multicolumn{1}{c}{\textbf{Threshold}} &
        \multicolumn{1}{c}{\textbf{$\mathbf{F1_{REAL}}$}} &
        \multicolumn{1}{c}{\textbf{$\mathbf{F1_{FAKE}}$}} &
        \multicolumn{1}{c}{\textbf{$\mathbf{ACC_{REAL}}$}} &
        \multicolumn{1}{c}{\textbf{$\mathbf{ACC_{FAKE}}$}} \\
        \midrule
        90 & 0.000049 & 0.9545 & 0.7125 & 0.9215 & 0.9215 \\
        92 & 0.000076 & 0.9466 & 0.6264 & 0.9065 & 0.9065 \\
        94 & 0.000109 & 0.9364 & 0.5064 & 0.8873 & 0.8873 \\
        96 & 0.000146 & 0.9260 & 0.3618 & 0.8674 & 0.8674 \\
        98 & 0.000197 & 0.9161 & 0.1916 & 0.8480 & 0.8480 \\
        \bottomrule
    \end{tabular}%
    }
\end{table}

\vspace{-0.09cm}

\section{Evaluation and Results}
\label{sec:eval-results}

\subsection{Implementation}

Experiments are run on the FABRIC testbed with compute nodes emulating SDN switches, edge servers, and a VR client. The anomaly detection dataset is generated by running ILLIXR in headless mode with Vulkan Swapchain disabled for CLI compatibility. We deploy Floodlight’s STATISTICS-based load balancer, polling per-port byte counts every 2s and routing flows to the switch with the lowest delta. The pool comprises four Open vSwitch instances, each linked to an edge server performing pose estimation, similar to \cite{remotevio}.


All components—controller, switches, edge servers, and traffic generator—run on separate FABRIC nodes (Fig.~\ref{fig:sdn-setup}). A fifth node generates the ILLIXR-driven VR workload and ICMP background traffic with exponential inter-arrival time. Each epoch is labeled \textit{REAL} or \textit{FAKE} based on misreporting configuration. The workload uses ILLIXR’s Offload VIO client, offloading pose data every 15s. Edge servers (2-core CPUs, 4\,GB RAM, 100\,GB disk, NVIDIA A30 GPUs) execute the VIO pipeline; switches use similar nodes without GPUs. The SDN controller (4-core CPU) runs Floodlight, coordinating flow decisions over FABRIC’s internal network.


\subsection{Experimental Configuration}

We collect about 48 hours of VR activity data on the FABRIC testbed using ILLIXR. Each session begins with normal load reporting, followed by a period where one switch performs stealthy misreporting. Four sessions are recorded with varying attack parameters \(\rho\), \( \tau \), and \(\varphi\), summarized in Table~\ref{tab:attack-parameters}.

\vspace{-0.1cm}

\begin{table}[htbp]
\centering
\caption{Attack parameter combinations}
\label{tab:attack-parameters}
\footnotesize
\begin{tabular}{
  S[table-format=1.2]
  S[table-format=1.2]
  S[table-format=1.2]
}
\toprule
{\boldmath$\rho$} & {\boldmath$\tau$} & {\boldmath$\varphi$} \\
\midrule
0.01 & 0.48 & 0.31 \\
0.01 & 0.29 & 0.06 \\
0.10 & 0.48 & 0.48 \\
0.10 & 0.29 & 0.09 \\
\bottomrule
\end{tabular}
\end{table}


Each load-balancing action is followed by a 15s window where the headset offloads pose estimation, so misreporting can influence server selection and QoE. The dataset is shuffled and split into training (60\%), validation (20\%), and testing (20\%), each with 1000 epochs. Misreporting is modeled via Bernoulli sampling to match \(\varphi\). Models are trained offline on a 24-core CPU, 32\,GB RAM, and NVIDIA RTX 2000 Ada GPU, but are lightweight enough for real-time deployment on the SDN controller.

\subsection{Results}
\subsubsection{Classifier Effectiveness}

The MLP and calibrated LightGBM achieve high accuracy ($0.9659$, $0.9661$) and AUC ($0.9845$, $0.9873$) as shown in Tab.~\ref{tab:clf-detailed}. FAKE detection is robust, with precision $[0.9178, 0.9198]$ and recall $[0.8779, 0.8768]$, while REAL precision and recall both exceed $0.97$. These results show that shallow classifiers, combined with unsupervised features (reconstruction error, Mahalanobis distance, statistical deviation), provide strong separability. The hybrid design enables accurate, low-latency misreporting detection for real-time SDN use.

\begin{table}[htb]
\centering
\caption{Test-set performance of MLP and LightGBM}
\label{tab:clf-detailed}
\setlength{\tabcolsep}{6pt}
\renewcommand{\arraystretch}{1.1}
\begin{tabular}{lcc}
\toprule
\textbf{Metric} & \textbf{MLP} & \textbf{LightGBM} \\
\midrule
REAL Precision     & 0.9753 & 0.9750 \\
REAL Recall        & 0.9839 & 0.9843 \\
REAL F1-score      & 0.9796 & 0.9796 \\
FAKE Precision     & 0.9178 & 0.9198 \\
FAKE Recall        & 0.8779 & 0.8768 \\
FAKE F1-score      & 0.8974 & 0.8978 \\
Accuracy           & 0.9659 & 0.9661 \\
ROC AUC            & 0.9845 & 0.9873 \\
\bottomrule
\end{tabular}
\end{table}

\begin{figure}[t]
    \centering
    \begin{subfigure}[t]{0.49\linewidth}
        \centering
        \includegraphics[width=\linewidth]{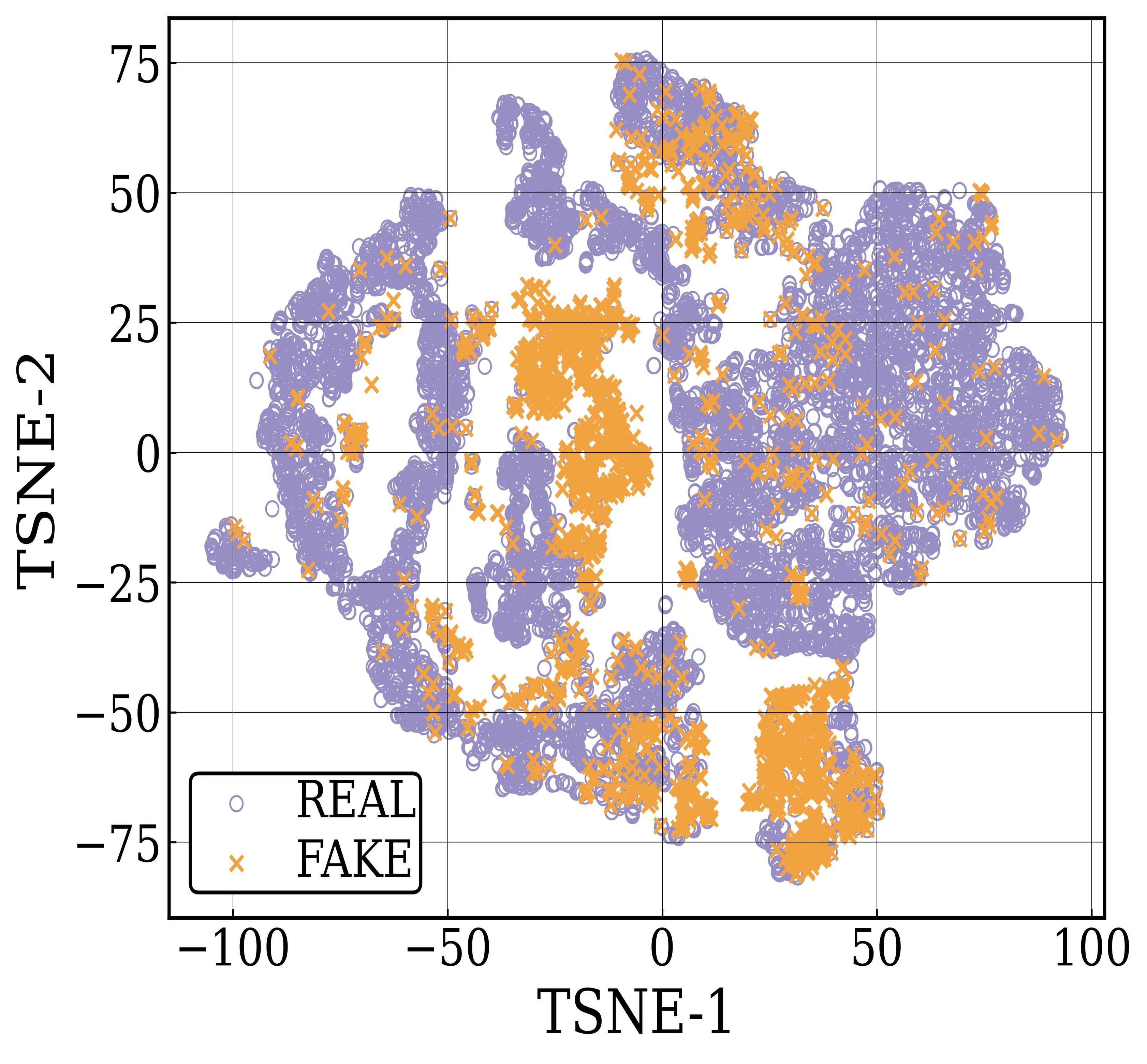}
        \caption{t-SNE: partial separation in latent space.}
        \label{fig:hybrid-tsne_latent_space}
    \end{subfigure}
    \hfill
    \begin{subfigure}[t]{0.49\linewidth}
        \centering
        \includegraphics[width=\linewidth]{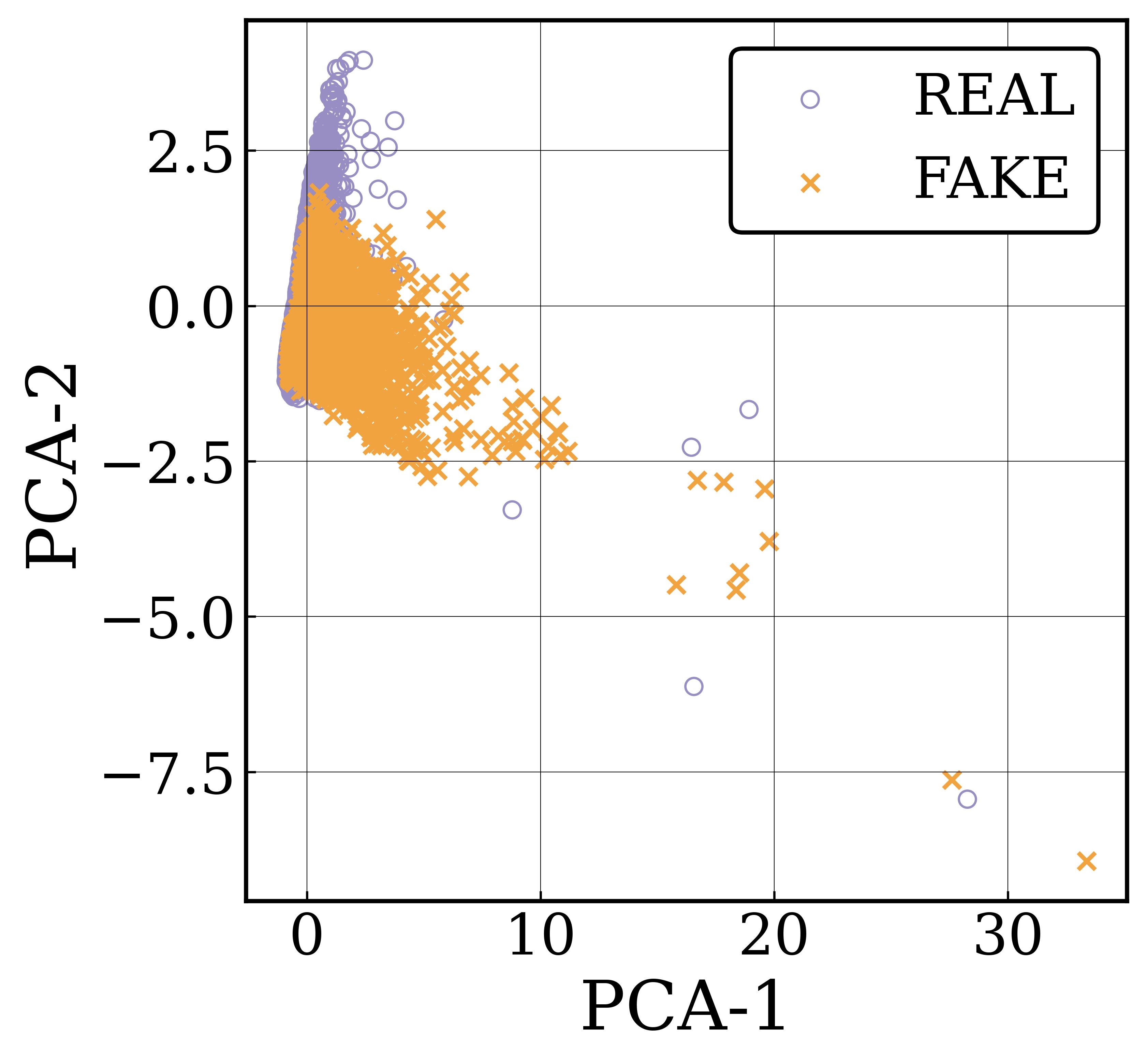}
        \caption{PCA: broader FAKE variance on PC1.}
        \label{fig:hybrid-pca_projection}
    \end{subfigure}
    \vspace{-1mm}
    \caption{Low-dimensional projections of latent and fused anomaly features.
    {\color{violet}Purple} = REAL, {\color{orange}Orange} = FAKE}
    \label{fig:hybrid-tsne-pca-combined}
    \vspace{-2mm}
\end{figure}

\subsubsection{Latent Embedding Analysis}

t-SNE projections of latent features show FAKE samples clustering more tightly and REAL more dispersed (Fig.~\ref{fig:hybrid-tsne_latent_space}), indicating that the autoencoder captures some class-relevant structure. However, substantial overlap shows latent embeddings alone lack clear separation. PCA on fused features reveals PC1 explains most variance, with FAKE spanning a wider range than REAL (Fig.~\ref{fig:hybrid-pca_projection}), highlighting greater variability in FAKE embeddings and their value for detection.

\subsubsection{Model Explainability via SHAP}

\begin{figure}[htbp]
    \centering
    \includegraphics[width=0.75\columnwidth]{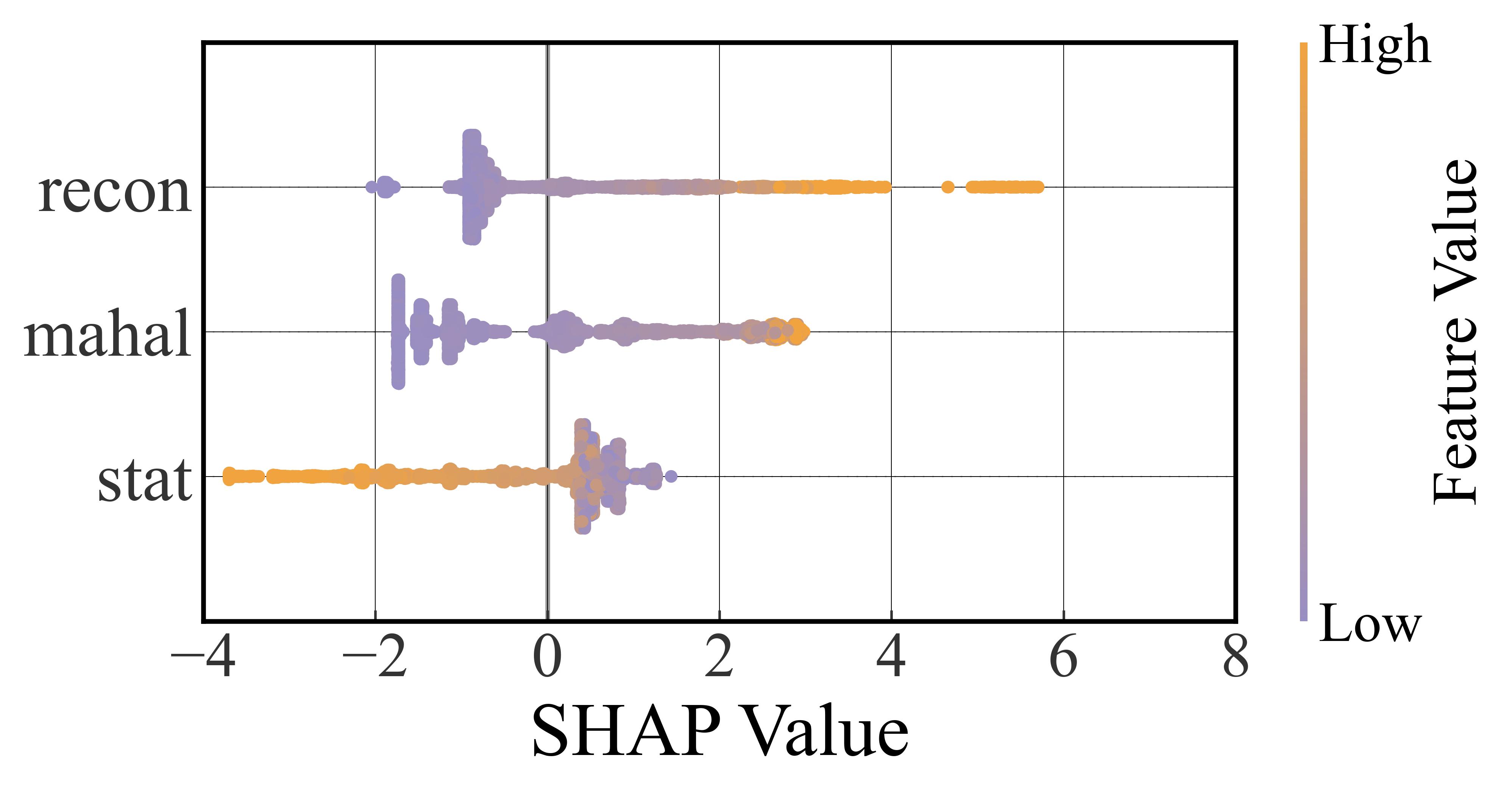}
    \caption{Global SHAP summary plot showing top contributing features towards classification.}
    \label{fig:hybrid-shap_summary_plot}
\end{figure}

We use SHAP (SHapley Additive exPlanations) to interpret contributions of the fused anomaly features. Fig.~\ref{fig:hybrid-shap_summary_plot} shows the global SHAP summary ranking features by impact on FAKE. Reconstruction error (\texttt{recon}) dominates, with high values strongly pushing predictions toward FAKE. Mahalanobis distance (\texttt{mahal}) provides moderate, consistent support, while statistical deviation (\texttt{stat}) has weaker, symmetric influence—high values often favor REAL, low values give mild FAKE support. Overall, \texttt{recon} leads, \texttt{mahal} supports, and \texttt{stat} plays a minor role.


\begin{figure}[htbp]
    \centering
    \includegraphics[width=\linewidth]{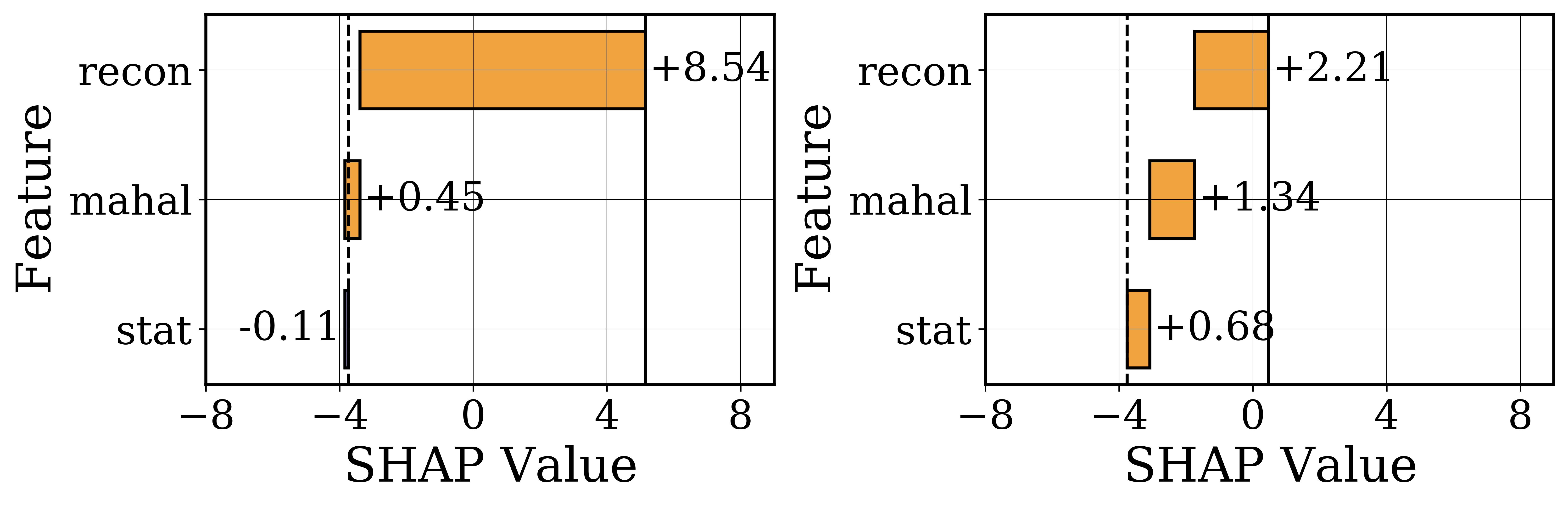}
    \vspace{-1mm}
    \caption{SHAP waterfall plots from the hybrid model explaining a true positive and false negative prediction.
    \textcolor{violet}{Purple} = REAL
    \textcolor{orange}{Orange} = FAKE. 
    Dashed = $E[f(x)]$, Solid = $f(x)$.}
    \label{fig:hybrid-shap-waterfall-combined}
\end{figure}

Instance-level attributions illustrate this further. In the true positive case (Fig.~\ref{fig:hybrid-shap-waterfall-combined}, left), strong \texttt{recon} (+8.54) with \texttt{mahal} (+0.45) outweighs weak negative \texttt{stat} ($-0.11$), pushing the logit above the FAKE threshold. In the false negative case (right), all features are positive—\texttt{recon} (+2.21), \texttt{mahal} (+1.34), \texttt{stat} (+0.68)—but the combined signal is insufficient to cross the boundary. This reflects a failure mode where anomaly cues exist but lack strength for detection.


\subsubsection{Comparison with Unsupervised Methods}

For comparison with a fully unsupervised variant, we replace supervised classifiers with Isolation Forest, One-Class SVM, GMM (Combined and Latent), LOF, and KMeans (Latent). These models use either fused anomaly scores (reconstruction error, statistical deviation, Mahalanobis distance) or Transformer latent features, ensuring fairness against the hybrid baseline.


Tab.~\ref{tab:unsupervised_fake_detection} summarizes results. Isolation Forest performs best among unsupervised models with $\mathrm{F1}_{\text{FAKE}}=0.6984$ and $\mathrm{AUC}=0.9339$, capturing high-dimensional deviations but with many false positives from uncalibrated thresholds. One-Class SVM maintains precision but recalls only $32.26\%$ of attacks. GMM (Combined) achieves recall $0.7776$ but low precision ($\mathrm{AUC}=0.6567$), while LOF and KMeans show very low $\mathrm{F1}$. Overall, Isolation Forest detects subtle anomalies but remains unreliable, whereas our Transformer-calibrated hybrid reaches $\mathrm{F1}_{\text{FAKE}}>0.80$, combining anomaly sensitivity with supervised calibration, interpretability, and real-time viability for SDN defense.


\begin{table}[t]
    \centering
    \footnotesize
    \setlength{\tabcolsep}{6pt} 
    \caption{Transformer autoencoder performance}
    \label{tab:transformer_thresholds}
    \resizebox{\linewidth}{!}{%
    \begin{tabular}{
        S[table-format=2.0]
        S[table-format=1.6]
        S[table-format=1.4]
        S[table-format=1.4]
        S[table-format=1.4]
        S[table-format=1.4]
    }
        \toprule
        \multicolumn{1}{c}{\textbf{Percentile}} &
        \multicolumn{1}{c}{\textbf{Threshold}} &
        \multicolumn{1}{c}{\textbf{$\mathbf{F1_{REAL}}$}} &
        \multicolumn{1}{c}{\textbf{$\mathbf{F1_{FAKE}}$}} &
        \multicolumn{1}{c}{\textbf{$\mathbf{ACC_{REAL}}$}} &
        \multicolumn{1}{c}{\textbf{$\mathbf{ACC_{FAKE}}$}} \\
        \midrule
        90 & 0.000049 & 0.9545 & 0.7125 & 0.9215 & 0.9215 \\
        92 & 0.000076 & 0.9466 & 0.6264 & 0.9065 & 0.9065 \\
        94 & 0.000109 & 0.9364 & 0.5064 & 0.8873 & 0.8873 \\
        96 & 0.000146 & 0.9260 & 0.3618 & 0.8674 & 0.8674 \\
        98 & 0.000197 & 0.9161 & 0.1916 & 0.8480 & 0.8480 \\
        \bottomrule
    \end{tabular}%
    }
\end{table}

\subsubsection{Deployment Cost and Inference Latency}

Tab.~\ref{tab:inference_time_comparison} shows that the MLP, statistical autoencoder, Mahalanobis scorer, and LightGBM all meet sub-millisecond budgets. The only exception is the Transformer autoencoder, which records 1.745,s on a resource-constrained FABRIC controller. In realistic deployments, modern edge controllers (e.g., Edgecore AS7326-56X with Intel Xeon D-1518 CPU and 16GB DDR4 RAM) are far more capable, so this latency would be greatly reduced.

\begin{table}[htbp]
\centering
\caption{Per-sample inference latency across modules}
\label{tab:inference_time_comparison}
\footnotesize
\begin{tabular}{l S[table-format=1.4]}
\toprule
\textbf{Model} & {\textbf{Time} (\si{\second})} \\
\midrule
MLP                     & 0.0010 \\
Statistical Autoencoder & 0.0050 \\
Mahalanobis             & 0.0147 \\
Calibrated LightGBM     & 0.0311 \\
Transformer Autoencoder & 1.7451 \\
\bottomrule
\end{tabular}
\end{table}

\subsubsection{Fusion Component Ablation Study}
We evaluate the contribution of reconstruction error, statistical deviation, and Mahalanobis distance, reported in Tab.~\ref{tab:ablation_results} using an ablation study. Using reconstruction alone gives the strongest signal ($\mathrm{F1}_{\text{FAKE}}=0.880$, $\mathrm{AUC}=0.980$), while Mahalanobis offers moderate performance ($\mathrm{F1}_{\text{FAKE}}\approx0.73$, $\mathrm{AUC}>0.94$). Statistical deviation collapses ($\mathrm{F1}_{\text{FAKE}}=0$, $\mathrm{AUC}\approx0.658$). Fusion of all three boosts LightGBM to $\mathrm{F1}_{\text{FAKE}}=0.898$, $\mathrm{AUC}=0.987$, with MLP trailing by $<0.02$. Across branches, LightGBM slightly outperforms MLP, especially under fusion.


\begin{table}[htbp]
\centering
\caption{Ablation study}
\label{tab:ablation_results}
\footnotesize
\resizebox{\linewidth}{!}{%
\begin{tabular}{
  l
  S[table-format=1.4]
  S[table-format=1.4]
  S[table-format=1.4]
  S[table-format=1.4]
}
\toprule
\textbf{Model} &
\multicolumn{1}{c}{\textbf{Precision\textsubscript{FAKE}}} &
\multicolumn{1}{c}{\textbf{Recall\textsubscript{FAKE}}} &
\multicolumn{1}{c}{\textbf{F1\textsubscript{FAKE}}} &
\multicolumn{1}{c}{\textbf{AUC}} \\
\midrule
MLP (Recon)            & 0.8936 & 0.8666 & 0.8799 & 0.9796 \\
LightGBM (Recon)       & 0.8988 & 0.8563 & 0.8771 & 0.9786 \\
MLP (Stat)             & 0.0000 & 0.0000 & 0.0000 & 0.6582 \\
LightGBM (Stat)        & 0.0000 & 0.0000 & 0.0000 & 0.6568 \\
MLP (Mahalanobis)      & 0.7274 & 0.7394 & 0.7334 & 0.9503 \\
LightGBM (Mahalanobis) & 0.7381 & 0.7166 & 0.7272 & 0.9488 \\
MLP (Fusion)           & 0.9177 & \bfseries 0.8779 & 0.8974 & 0.9845 \\
LightGBM (Fusion)      & \bfseries 0.9198 & 0.8768 & \bfseries 0.8978 & \bfseries 0.9873 \\
\bottomrule
\end{tabular}%
}
\end{table}






\subsubsection{Cross-Dataset Generalization Benchmark}

Tab.~\ref{tab:benchmark_cross-dataset} benchmarks cross-dataset generalization. 
With longer session intervals, performance drops sharply: in \texttt{interval\_25}, 
FAKE F1 drops to 0.21, REAL F1 to 0.17, and AUC to 0.45. 
\texttt{interval\_20} recovers REAL F1 (0.79) but leaves FAKE F1 low (0.24) 
due to class imbalance and attack stealth. In contrast, \texttt{epoch\_500} 
remains strong (FAKE F1 = 0.65, REAL F1 = 0.96, AUC = 0.90), 
showing robustness to attack window variation but sensitivity to session timing.


\begin{table}[htbp]
\centering
\caption{Cross-dataset benchmarking metrics}
\label{tab:benchmark_cross-dataset}
\footnotesize
\resizebox{\linewidth}{!}{%
\begin{tabular}{
  l
  S[table-format=1.6]
  S[table-format=1.6]
  S[table-format=1.6]
  S[table-format=1.6]
}
\toprule
\textbf{Dataset} &
\multicolumn{1}{c}{\textbf{F1\textsubscript{FAKE}}} &
\multicolumn{1}{c}{\textbf{F1\textsubscript{REAL}}} &
\multicolumn{1}{c}{\textbf{AUC}} &
\multicolumn{1}{c}{\textbf{Accuracy}} \\
\midrule
Dataset\_interval\_25 & 0.206571 & 0.174460 & 0.451225 & 0.190834 \\
Dataset\_interval\_20 & 0.243271 & 0.787129 & 0.562899 & 0.667727 \\
Dataset\_epoch\_500   & 0.646173 & 0.957318 & 0.896945 & 0.923825 \\
\bottomrule
\end{tabular}%
}
\end{table}


\subsubsection{Sliding Window Sensitivity - Window vs. Stride}


Finally, Tab.~\ref{tab:benchmark_window_stride} reports sensitivity to window and stride. Larger windows (15–20) with moderate strides (5–10) give the best results, with FAKE F1 $>0.91$ and REAL F1 $>0.98$ under settings like (20, 10). LightGBM consistently edges out MLP in AUC and accuracy. Stride = 15 causes slight degradation from under-sampling, confirming that over-aggregation reduces anomaly sensitivity. Overall, performance depends strongly on temporal context length and overlap. All evaluation related codes and data are available through Github~\cite{git-xr-cnsm}.

\begin{table}[htbp]
\centering
\caption{Sliding-window sensitivity: MLP vs.\ LightGBM}
\label{tab:benchmark_window_stride}
\footnotesize
\resizebox{\linewidth}{!}{%
\begin{tabular}{
  S[table-format=2.0]
  S[table-format=2.0]
  S[table-format=1.3]
  S[table-format=1.3]
  S[table-format=1.3]
  S[table-format=1.3]
  S[table-format=1.3]
  S[table-format=1.3]
}
\toprule
 &  & \multicolumn{2}{c}{\textbf{F1\textsubscript{FAKE}}}
    & \multicolumn{2}{c}{\textbf{F1\textsubscript{REAL}}}
    & \multicolumn{2}{c}{\textbf{AUC}} \\
\cmidrule(lr){3-4}\cmidrule(lr){5-6}\cmidrule(lr){7-8}
\textbf{Window} & \textbf{Stride} & \textbf{MLP} & \textbf{LGB} & \textbf{MLP} & \textbf{LGB} & \textbf{MLP} & \textbf{LGB} \\
\midrule
5  & 5  & 0.774 & 0.751 & 0.963 & 0.962 & 0.960 & 0.961 \\
10 & 5  & 0.887 & 0.884 & 0.978 & 0.977 & 0.978 & 0.978 \\
10 & 10 & 0.833 & 0.830 & 0.965 & 0.966 & 0.971 & 0.971 \\
15 & 5  & 0.892 & 0.899 & 0.978 & 0.979 & 0.981 & 0.981 \\
15 & 10 & 0.867 & 0.875 & 0.972 & 0.974 & 0.978 & 0.979 \\
15 & 15 & 0.830 & 0.847 & 0.965 & 0.969 & 0.963 & 0.963 \\
20 & 10 & 0.911 & 0.916 & 0.981 & 0.982 & 0.986 & 0.986 \\
20 & 15 & 0.885 & 0.883 & 0.975 & 0.975 & 0.982 & 0.982 \\
\bottomrule
\end{tabular}%
}
\end{table}


\section{Conclusions and Future Work}
\label{sec:Conclusion}

This paper examined stealthy misreporting attacks in SDN-hosted VR systems. Using the FABRIC testbed and ILLIXR framework, we built a realistic SDN-VR pipeline, reproduced misreporting, and demonstrated its impact on pose tracking and scene stability. To counter this, we proposed a hybrid detection framework that combines a Transformer-based autoencoder with statistical and latent-space features, fused into a lightweight supervised classifier, improving F1 by up to 20\% over baselines. 
Future work will explore coordinated multi-switch attacks and application-layer feedback to further strengthen detection in immersive SDN environments.


\bibliographystyle{ieeetr}
\bibliography{refs}

\end{document}